\documentclass[12pt]{article}
\usepackage{hyperref}
\usepackage{cite}
\usepackage{color}
\usepackage{graphicx}
\usepackage{amsmath}
\usepackage{amssymb}
\usepackage{xspace}

\makeatletter
\@addtoreset{equation}{section}

\makeatletter
\renewcommand\section{\@startsection {section}{1}{\z@}%
                                   {-3.5ex \@plus -1ex \@minus -.2ex}
                                   {2.3ex \@plus.2ex}%
                                   {\normalfont\large\bfseries}}
\renewcommand\subsection{\@startsection{subsection}{2}{\z@}%
                                     {-3.25ex\@plus -1ex \@minus -.2ex}%
                                     {1.5ex \@plus .2ex}%
                                     {\normalfont\bfseries}}

\def\baselinestretch{1.2}
\newcommand{\be}{\begin{equation}}
\newcommand{\ee}{\end{equation}}
\newcommand{\beq}{\begin{eqnarray}}
\newcommand{\eeq}{\end{eqnarray}}

\newcommand{\gone}[1]{{}}

\begin{document}
\begin{titlepage}
\begin{flushright}
MAD-TH-15-05
\end{flushright}

\vfil

\begin{center}

{\bf \Large
On Axionic Field Ranges, Loopholes and the Weak Gravity Conjecture}

\vfil

Jon Brown, William Cottrell, Gary Shiu and Pablo Soler

\vfil

Department of Physics, University of Wisconsin, Madison, WI 53706, USA\\

\vfil

\end{center}

\begin{abstract}
\noindent In this short note we clarify some aspects of the impact that the Weak Gravity Conjecture has on models of (generalized) natural inflation. We address in particular certain technical and conceptual concerns recently raised regarding the stringent constraints and conclusions found in our previous work~\cite{Brown:2015iha}. We also point out the difficulties faced by attempts to evade these constraints. These new considerations improve the understanding of the quantum gravity constraints we found and further support the conclusion that it remains challenging for axions  to drive natural inflation.

\end{abstract}
\vspace{0.5in}

\end{titlepage}
\renewcommand{\baselinestretch}{1.05}  

\section{Introduction}

It is generally believed that understanding the nature of the big bang requires insights from quantum gravity. An appealing feature of inflation is that one can
temporarily sidestep the unknown mysterious ultraviolet (UV) completion of general relativity in drawing observational consequences of the early universe.
Yet, not all UV sensitivity is lost as Planck suppressed corrections tend to drive the inflaton mass to values larger than the Hubble scale during inflation.
Large field inflation highlights this sensitivity to Planck scale physics even further. As the inflaton traverses a super-Planckian distance in field space, its potential is subject to {\it all} Planck suppressed corrections, not only the leading ones. It is therefore natural to ask whether some general properties of quantum gravity can provide a guiding principle for inflationary scenarios of this type.

Some folk properties of quantum gravity, based on black hole physics, are succinctly summarized in the so-called ``weak gravity conjecture" (WGC)\cite{ArkaniHamed:2006dz, Ooguri:2006in}.
This conjecture applies readily to particles and gauge fields, but the generalization to other p-form fields is less apparent. 
Such generalizations are important since axions, which arise from higher-form fields in string theory, are popular inflaton candidates for large field inflation. 
The WGC for p-forms was recently studied and some initial steps in formulating the criteria for large field inflation were addressed
in \cite{Brown:2015iha}. 
Interestingly, the criteria on the axion field range agree with those obtained by analyzing the gravitational instantons to the axion potential
\cite{Montero:2015ofa} and general expectations on the axion moduli space \cite{Rudelius:2015xta}.

The quantum gravity constraints on large field inflation recently found \cite{Brown:2015iha,Montero:2015ofa,Rudelius:2015xta} are most apparent in natural inflation \cite{Freese:1990rb}. The WGC summarizes the generic problem one faces: the  flatness of the inflaton potential
requires the axion decay constant to be super-Planckian which is precisely when higher order instanton corrections become important. 
The bound in \cite{Brown:2015iha} precisifies  the domain of analytic control.
To get around this problem, several inflationary scenarios with multiple axions have been proposed \cite{Kim:2004rp, Dimopoulos:2005ac, Choi:2014rja, Long:2014dta, Shiu:2015uva, Shiu:2015xda, Bachlechner:2014gfa}. Nonetheless, the convex hull condition stated in \cite{Cheung:2014vva,Brown:2015iha} enables one to diagnose whether such more general models can evade the bound.
Except for models with monodromy \cite{McAllister:2008hb,Silverstein:2008sg,Marchesano:2014mla,Blumenhagen:2014gta,Hebecker:2014eua}, many of the axion inflation models are severely constrained by the weak gravity conjecture and appear to reside in the ``swampland" \cite{Vafa:2005ui}. 

Unlike the single axion case, defining the maximum field range for multiple axions is  subtle. Several proxies of the axion field range have been suggested in the literature. 
In \cite{Brown:2015iha} we presented a bound on one such proxy. 
Here, we show that the criteria stated in \cite{Brown:2015iha} also bounds the ``diameter" defined in 
\cite{Bachlechner:2015qja}. It is conceivable that the same methodology will apply to other proxies as well.

As with other ``no-go theorems", a clearly stated result also makes clear
   the possible loopholes. Indeed,
an idea to evade the stringent bound on the axion field range,  (which would be plausible only if the strong form of the weak gravity conjecture is violated), 
has been suggested in \cite{Brown:2015iha,Montero:2015ofa}.
This suggestion has recently been taken up by \cite{Bachlechner:2015qja, Hebecker:2015rya}. We will elaborate further on this loophole and point out a number of hurdles one needs to overcome.

This paper is organized as follows. In Section 2 we prove that the strong form of the weak gravity conjecture puts a bound on the diameter of axion field space and thus, if true, would prevent many models of natural inflation from being realized.  In Section 3 we discuss what it would entail to evade the strong form of the weak gravity conjecture via the loophole suggested in \cite{Brown:2015iha}.  Finally, we offer our conclusions.  

\section{Technical issues: the axion ``decay constants''}
We will consider a Lagrangian of the following form:\footnote{We assume that $M_{p}=1$ for simplicity}
\beq
\label{genlag}
\mathcal{L}=-\frac{1}{2}\sum_{i=1}^{N}\partial_{\mu} c^{i} \partial^{\mu}c^{i}-\Lambda^{4}\sum_{k=1}^{M} e^{-m_{k}} \left(1-\cos\left(\vec{\mathcal{Q}}_{k}\cdot \vec{c}\right)\right)
\eeq
We have chosen to canonically normalize the kinetic terms at the expense of having non-integral charges $\mathcal{\vec{Q}}_{k}$. The vector indices on $\vec{c}$, $\vec{\mathcal{Q}}_{k}$ run over all axions, of which there are $N$, while the index $k$ runs over all $M$ instantons.  We assume that $M\ge N$ in order to guarantee the eventual stabilization of all axions. 

Now, a general Lagrangian of the form (\ref{genlag}) may be quite complicated and the notion of an `axion decay constant' is somewhat ambiguous.  It is nevertheless useful to have a quick diagnostic tool in order to determine at a glance when inflation may occur.  In \cite{Brown:2015iha} the authors studied the eigenvalues of $\mathcal{Q}^{\top}\mathcal{Q}$ as a proxy for field range.  Moreover, it was demonstrated that whenever the WGC  is satisfied, the eigenvalues of this matrix (roughly corresponding to the inverse decay constants squared) will be larger than some fixed $\mathcal{O}(1)$ constant.  However, it has been suggested that a better measure for the possibility of inflation is just the net field range as constrained by the periodicity of the cosine terms.    This is the field diameter, ``$\mathcal{D}$" defined in \cite{Bachlechner:2014gfa}.  In this section, we will show that this parameter is also constrained by the WGC to be less than $2\pi$ if one wishes to stay within the regime of perturbative control.\footnote{There is technically an order 1 constant, $r_{(p,d)}$ discussed in \cite{Brown:2015iha} which defines what we mean by `perturbative control'.  Here, we will ignore this subtlety since it will not affect our main points.}   The proof presented here is similar to the one given in \cite{Brown:2015iha},  the proxy that is targeted is different.  

The diameter, $\mathcal{D}$, is defined as follows.  First, pick a unit normalized vector $\vec{c}_{0}$.  This vector defines a line in field space which we will parameterize as $s \vec{c}_{0}$.  We may associate a diameter with this line by constraining the argument of each cosine to lie between $-\pi$ and $\pi$.  In other words, we must require $s$ to satisfy :
\be
-\pi \le s \vec{\mathcal{Q}}_{k}\cdot \vec{c}_{0} \le \pi \,\,\,\,\,\,\,\,\,\,\,\,\,\,\,\,\,\,\,\,\,\,\,\, \forall k
\ee
The strongest constraint obviously comes from the largest $\vec{\mathcal{Q}}_{k}\cdot \vec{c}_{0}$.  We may thus define a diameter for this particular line as:
\be
\label{Dcdef}
\mathcal{D}(\vec{c}_{0}) = \frac{2\pi}{\max_{k} (\vec{\mathcal{Q}}_{k}\cdot \vec{c}_{0})}
\ee
The largest possible diameter is then written as $\mathcal{D} = \sup_{\{\vec{c}_{0}\}} \mathcal{D}(\vec{c}_{0})$.    

We now show that the WGC implies that $\forall \vec{c}_{0},\, \mathcal{D}(\vec{c}_{0}) \le 2\pi$ if one is in the regime of perturbative control.   To see this, recall that the WGC implies that the following vectors form a convex hull that contains the unit ball:
\be
\vec{z}_{k}=\frac{\vec{\mathcal{Q}}_{k}}{m_{k}}
\ee
This is equivalent to the statement that any normalized vector $\vec{c}_{0}$ may be written as:
\be
\label{chdef}
\rho \vec{c}_{0} = \sum_{k} \alpha_{k} \vec{z}_{k} = \sum_{k} \frac{\alpha_{k}}{m_{k}} \vec{\mathcal{Q}}_{k}
\ee
where $\rho \ge 1$ and $\sum_{k}\alpha_{k} =1$.  Now, suppose on the contrary that $\mathcal{D}(\vec{c}_{0}) > 2\pi$.  Equation (\ref{Dcdef}) would then imply:
\be
\label{largerange}
\vec{\mathcal{Q}}_{k}\cdot \vec{c}_{0} < 1,\,\,\,\,\,\,\,\,\,\,\,\,\,\,\,\,\,\,\,\,\,\,\,\, \forall k
\ee
If we define $q_{ij} \equiv \vec{\mathcal{Q}}_{i}\cdot \vec{\mathcal{Q}}_{j}$, then the equation above may be written more conveniently as:
\be
\label{largerange2}
\sum_{j} \frac{\alpha_{j}}{m_{j}}\, q_{jk} < \rho
\ee
Now, square (\ref{chdef}) and then plug in equation (\ref{largerange2}) :
\beq
\rho^{2}&=&\sum_{ij} \frac{\alpha_{i}\alpha_{j}}{m_{i}m_{j}} q_{ij} \\ \nonumber
&=&\sum_{i} \frac{\alpha_{i}}{m_{i}} \left(\sum_{j}\frac{\alpha_{j}}{m_{j}}q_{ij}\right) \\ \nonumber
&<& \rho \sum_{i} \frac{\alpha_{i}}{m_{i}} \\ \nonumber
&<&\rho
\eeq
In the last line we have used $\sum_{k}\alpha_{k} =1$ from the convex hull condition and also $m_{k}>1$ since we are in the regime of perturbative control.  We have thus derived $\rho^{2}< \rho$, in contradiction with the convex hull condition which requires $\rho \ge 1$.  We therefore conclude that constructing a well-controlled model with diameters larger than $2\pi$ is inconsistent with the WGC.  

\section{Conceptual issues: attempts to evade the constraints}
We have seen that a potential generated by instantons that satisfy the WGC is likely not suitable for inflation, since the range over which this potential is smooth enough is always sub-Planckian. The only way to evade this problem is to assume that these instantons (the ones that ensure that the WGC is satisfied) give negligible contributions and, on top of them, some other dominant non-perturbative effects generate a long range potential.  As will be detailed later, this condition satisfies only the mild-WGC and violates the strong-WGC.

The idea can be exemplified already in the case with a single axion, as already described in~\cite{Brown:2015iha}. Let us assume that the potential for such an axion receives two types of contributions, which for the canonically normalized axion $c$ reads:
\be
V= \Lambda_1^4 \, e^{-m}\left[1-\cos\left(\frac{c}{f}\right)\right]+ \Lambda_2^4 \, e^{-M}\left[1-\cos\left(\frac{k\,c}{f}\right)\right]\,,
\ee
where $k\in\mathbb{Z}$. The first term could lead to a suitable inflationary potential if $f>M_p$, while remaining under perturbative control would require $m>1$. Since these terms have $fm>M_p$, one would need to use the second term in the potential to satisfy the WGC, hence requiring $\frac{fM}{k}<M_p$. Finally, the latter instanton corrections  should be negligible in comparison with respect to the former, which would require $m<M$ (assuming that the coefficients $\Lambda_1\sim \Lambda_2$ are of the same order). These conditions are not easy to satisfy, especially taking into account that one cannot make $k$ parametrically large if one wants to avoid conflicts with the number of species problem and the Covariant Entropy Bound.

Another funny property of the scenario described above is that (in the particle picture) the more massive state $(M)$ is below the extremality bound, while the lightest state $(m)$ is above it. To be clear, the extremality bound mentioned here is the one that applies to macroscopic black holes; microscopic particles can consistently sink below the bound.  However, going below the bound is a purely quantum property (essentially it is the Casimir energy in perturbative string examples), and is hence expected to have a stronger effect on light states than in heavy ones.  For example, we consider $SO(32)$ heterotic on $T^{6}$, as discussed in cite(vafa).  In our conventions, the spectrum is:
\be
M=\sqrt{2}\left(Q_{L}^{2}+2N_{L}-2\right)^{1/2}
\ee

One must set $N_{L}=0$ in order to find a state which satisfies the convex hull condition.  Moreover, when $N_{L}=0$, one sees that $Q_{L}/M$ is monotonically decreasing as a function of $M$.  In other words, the generators of the convex hull are most likely to be the particles with the smallest mass while realizing the suggested loophole would require just the opposite.  Such a phenomenon has never been observed in a controlled setting.  Moreover, one should also worry that if the particles satisfying the WGC were very massive, as suggested by the loophole, then they would be within the regime of sugra and thus describe unphysical naked singularities.  

It therefore seems very plausible that the strong form of the WGC is satisfied and in this case the potential loophole may be excluded.  If the strong-WGC holds, the instantons with lowest action (the main contributions to the potential) satisfy the WGC and lead to sub-Planckian field ranges. As mentioned already in~\cite{ArkaniHamed:2006dz}, there is no known example in string theory where the strong-WGC is violated.

Let us nevertheless suppose that these challenges can be met and consider the generalization to cases with more axions where new ingredients can be used.  In particular, we focus on axion alignment. The general conclusion is nevertheless unchanged: in order to obtain a suitable inflation potential one needs a set of negligible ``spectator'' instantons that ensure that the WGC is satisfied and, on top of them, a set of dominant non-perturbative contributions that generate a potential smooth enough along a large field range. This field range cannot be made parametrically large without encountering a species problem. 

We will address in the following certain ideas that have been brought up recently which can be related to the use of this loophole to generate large field inflation. Hopefully, our discussion will clarify a few confusing points, and point out important difficulties that these methods have to face to achieve their goals.

\subsection{Gravitational instantons}
The authors of~\cite{Montero:2015ofa} took a bottom-up approach to study quantum gravitational effects on effective models  of natural inflation. They studied a low energy field theory consisting solely of $N$ axion fields $\phi^i$ coupled to gravity and found a set of instantonic solutions (referred to as ``gravitational instantons'') of the equations of motion. The effective Lagrangian of such a theory reads 
\begin{equation}\label{lag}
{\cal L}=\frac{1}{2}\left(\,\partial\vec{\phi}\,\right)^{\,T}\cdot {\cal G}\cdot \left(\,\partial\vec{\phi}\,\right)-\sum_{\vec{n}\in\mathbb{Z}^N} \Lambda^4 \,e^{-S_{\vec{n}}}\left[1-\cos\left(\vec{n}^{\,T}\cdot\vec{\phi}\right)\right]
\end{equation}
This Lagrangian is written in a basis such that the periodicity of the axions is $\phi^i\equiv \phi^i+2\pi$ while the kinetic term $\cal G$ can be non-diagonal.   

As described in~\cite{Montero:2015ofa}, for each vector $\vec{n}\in\mathbb{Z}^N$, there exists a gravitational instanton with classical action
\begin{equation}
S_{\vec{n}}^{(0)}=M\sqrt{\vec{n}^{\,T}\cdot {\cal G}^{-1}\cdot\vec{n}}
\end{equation}
with $M=\frac{\sqrt{3\pi}}{8}$.\footnote{A different value $M=\frac{\sqrt{6}\pi}{8}$ is obtained by~\cite{Bachlechner:2015qja}. For our following arguments, the precise value will not play an important role.} Such instantons generically generate potentials with sub-Planckian ranges, a result which is in line with the WGC\cite{Montero:2015ofa}. However, we stress that these instanton solutions were found in a low energy approximation, and cannot be trusted microscopically, i.e. for low values of $|\vec{n}|<|\vec{n}|_{min}$.  Moreover, the small values of $\vec{n}$ are precisely the ones which are important for generating the convex hull, since these are the ones for which the quantum corrections are most relevant and so it is not clear how much can be learned from a pure gravitational analysis.  

Nevertheless, let us consider the approach taken in~\cite{Montero:2015ofa}, and later also advocated in~\cite{Bachlechner:2015qja}, in which kinetic mixing putatively allows one to make the contributions of these gravitational instantons very small and, on top of them, could allow for some sizable extra non-perturbative effects whose contributions are dominant and generate a long range potential. The Lagrangian for such a theory would look like
\begin{equation}\label{lag2}
{\cal L}=\frac{1}{2}\left(\,\partial\vec{\phi}\,\right)^{\,T}\cdot {\cal G}\cdot \left(\,\partial\vec{\phi}\,\right)-\sum_{\vec{n}\in\mathbb{Z}^N} \Lambda^4 \,e^{-S_{\vec{n}}}\left[1-\cos\left(\vec{n}^{\,T}\cdot\vec{\phi}\right)\right]-\sum_k \tilde{\Lambda}^4\,e^{-\tilde{S}_k}\left[1-\cos\left(\vec{n}_k^{\,T}\cdot\vec{\phi}\right)\right]
\end{equation}
where on top of the gravitational instantons that satisfy the WGC, there are some extra contributions labeled by an index $k$ that generate the inflaton potential (and hence by themselves do not satisfy the WGC).\footnote{It is important to notice that the extra contributions must come from a different type ({\it family}) of objects other than the instantons that satisfy the WGC:  `bound states' (i.e. higher harmonics) of elementary instantons above the extremality bound are not expected to go below it.}
This setup represents precisely the loophole that was mentioned before. We will address this possibility in the following and will  point out several severe difficulties in its realization.

\subsection{Extra contributions}

The first question one needs to ask when one has identified the objects that satisfy the WGC and wants to invoke the loophole to generate large field inflation is: what are the extra non-perturbative contributions that generate the dominant potential? From the low energy perspective of the previous subsection, the origin of such new effects is mysterious, since in an effective theory consisting purely of axions and gravity, the only non-perturbative effects we know about are the described ``gravitational instantons''. In order to find such extra contributions, one would need to add extra ingredients to the low energy theory. One possibility would be to include gauge fields and hence ``gauge instantons'' (e.g. gaugino condensation) in the theory. 

However, one has to take into account that the ``gravitational instanton'' solutions were found in the absence of these new ingredients.  When the latter are taken into account (e.g. when gauge fields are added), the ``gravitational solutions'' would be modified, and it is important to address the consequences. In fact, it is not straightforward to assume that ``gravitational instantons'' and ``gauge instantons'' are different things that coexist simultaneously. In particular, this dichotomy seems to implicitly assume that ``gauge instantons" somehow do not couple to gravity.  One would need to find two different sets (``gravitational'' and ``gauge'') of back-reacted instanton solutions to the equations of motion, and study their properties. 

The picture is perhaps clearer from a top-down string perspective. A useful setup is that of type II string theory with  axions that arise from the reduction of RR $p$-forms along $p$ dimensional cycles. We will focus for concreteness on the case $p=4$. In the absence of gauge branes (and under the assumption that the {\it dilatonic} partners of these axions are stabilized at a high scale),\footnote{In fact, the presence of light dilatonic partners of the axions could have a very important effect on the ``gravitational instanton'' solutions, \cite{Bergshoeff:2005zf}.} the low energy theory contains just axions coupled to gravity, and hence the ``gravitational instantons'' described previously should be present. From the string perspective, the instanton effects that generate a potential for these RR axions are given by Euclidean D3-branes wrapping the 4-cycles. Hence, it is natural to identify the low energy ``gravitational instantons'' with the stringy Euclidean branes, as was argued already in~\cite{Montero:2015ofa}. It should not be surprising that these instantons satisfy the WGC since, in the absence of other ingredients, they are the only effects (that we know of) that generate a potential for the axions.

Now the idea to generate large field inflation is to add some extra ingredient to the setup, say gauge fields. This can be achieved by wrapping some of the $4$-cycles with D7-branes. Of course, the presence of the gauge branes would have a strong impact on the instantonic Euclidean D3-branes that generated the low-energy ``gravitational instantons''. In fact, 
gaugino condensation on a D7-brane can be interpreted as arising from ``fractional'' euclidean D3-branes wrapping the same cycle as the D7-branes. The picture is even more clear in the lift to F/M-theory, where both gaugino condensation and Euclidean D3-branes are represented by M5-branes wrapping the 4-cycles and the fibered torus. Hence, it seems misleading to consider ``gravitational instantons'' and ``gauge instantons'' as different objects that contribute simultaneously to the potential of the same axions. As the simplest example, if there is a single 4-cycle in the system, one would consider ``gravitational instantons''/Euclidean D3-branes if the cycle is not wrapped by a D7, OR ``gauge instantons''/``fractional'' D3-branes if the cycle is wrapped by a D7; but not both simultaneously.\footnote{Of course, we are not claiming that there will be a single instanton contribution for each axion. As in~\eqref{lag}, one would generically have contributions along each direction $\vec{n}$ in `charge' space, i.e. higher harmonics (they could be interpreted as `bound states' of elementary instantons, particles in a dual language). What one needs in order to use the loophole is a different {\it family} of instantons besides the one that satisfies the WGC.}  In this case the axion associated with a cycle wrapped by a gauge brane receives only contributions from ``gauge instantons'',  these would need to satisfy the WGC by themselves, and hence would not be useful in generating a large field potential.\footnote{Technically, the map from instantons to particles used in~\cite{Brown:2015iha} was described  in the absence of gauge branes. These would map to domain walls. We do not see an immediate obstacle in applying the arguments that lead to the WGC in the presence of these domain walls, since black holes could be studied arbitrarily far from the walls. This is nevertheless an important issue that would deserve further study.}

So our first conclusion is that in order to use the loophole to generate large field inflation in a complete setup, one needs to carefully address the origin of the different families of instantons: the spectator instantons that satisfy the WGC, and the dominant instantons that generate the inflationary potential. This is a crucial first step that has not been carefully discussed so far (with the exception of~\cite{Hebecker:2015rya}, on which we will briefly comment later).

\subsection{Violating the strong-WGC}

Once these instantons are identified, one would need to address the questions of why the spectator instantons are highly suppressed with respect to the dominant instantons, and why would the latter generate a large field range, i.e. why would the system violate the strong-WGC. 

The authors of~\cite{Montero:2015ofa} argued that, by kinetically aligning a set of $N$ axions, one could reach regimes in which the smaller gravitational instanton actions would be enhanced by a factor of $\sqrt{N}$ and hence the gravitational instanton contributions would be highly suppressed for large $N$. This suppression of the ``instanton number'' $\|\vec{n}\|\equiv\sqrt{\vec{n}^{\,T}\cdot {\cal G}^{-1}\cdot\vec{n}}$ was later shown to be statistically generic in the large $N$ limit in~\cite{Bachlechner:2015qja}. 

This, however, does not imply that the gravitational instantons would be suppressed with respect to the extra instantons (to which we will refer as gauge instantons despite the concerns raised in the previous subsection). Kinetic alignment affects equally all types of instantons: a generic instanton contribution along the direction $\vec{n}$ should be suppressed by the invariant length in charge space given by $\|\vec{n}\|$.

The classical actions for gravitational instantons and gauge instantons of the inflationary Lagrangian~\eqref{lag2} read, respectively,
\begin{equation}
S_{\vec{n}}^{(0)}=M \,\|\vec{n}\|\,\,,\qquad\qquad \tilde{S}_{\vec{n}_k}^{(0)}=\tilde{M} \, \|\vec{n}_k\|
\end{equation}
As mentioned before, $\vec{n}$ runs over all possible integer $N$-vectors, while $\vec{n}_k$ should only point in certain directions, so that a large field inflaton potential is indeed generated. In the optimal case, $\vec{n}_k$ would be aligned with the direction of shortest length in the axion domain, that is $\left(\vec{n}_k\right)^i=\delta_k^i$, ($k=1,\ldots,N$), so that the dominant potential along the diagonal direction is as flat as possible. 

One has to make sure that gravitational instanton contributions are negligible with respect to those from gauge instantons\footnote{We focus here on hierarchies generated by the classical instanton actions. Other perhaps more contrived options would be to tune the quantum corrections of the instanton actions, or the prefactor scales $\Lambda$ and $\tilde{\Lambda}$ at which the corrections appear.}
\begin{equation}\label{tune}
 \tilde{S}_{\vec{n}_k}^{(0)}< S_{\vec{n}}^{(0)}\qquad\Longrightarrow \qquad \tilde{M}< M\,\frac{\|\vec{n}\|}{\|\vec{n}_k\|}
\end{equation}
this hierarchy should hold for every $k$ and for every $\vec{n}$ that contributes to the potential in the inflationary (diagonal) direction. The fraction in the right hand side can always be made equal to one by considering the case $\vec{n}=\vec{n}_k$. It is hence a necessary condition that $\tilde{M} < M$. 
However, this  is not always sufficient. In particular if there is some $\vec{n}$ for which $\|\vec{n}\|<\|\vec{n}_k\|$, condition~\eqref{tune} becomes stronger. 

In fact, in the scenarios described in~\cite{Montero:2015ofa}, the kinetic matrix $\cal G$ was chosen so that the vector along the diagonal direction $\vec{n}_{diag}=(1,1,\ldots,1)$ had a much smaller length than any other vector, and in particular, $\|\vec{n}_{diag}\| \ll \|\vec{n_k}\|$. In~\cite{Bachlechner:2015qja} it was shown that this is actually a generic situation, and that instanton contributions along diagonal vectors, are the least suppressed ones in models with large $N$ and arbitrary kinetic mixing. As we see from eq.~\eqref{tune}, rather than an advantage, this is a drawback when trying to generate large field inflation while satisfying the mild-WGC, since the hierarchy between gauge instantons and spectator instantons needs to be stronger
\begin{equation}
\tilde{M}< M\,\frac{\|\vec{n}_{diag}\|}{\|\vec{n}_k\|}\ll M
\end{equation}
This is a first reason why it is more favorable to consider kinetic matrices such that vectors along the diagonal direction in axion space are more suppressed than the rest.

We have described here how would one need to suppress the spectator ``gravitational'' instantons with respect to the dominant ``gauge'' instantons in order to have large field inflation. In a complete setup, after the first step of identifying what each family of instantons corresponds to, one should make sure that the condition~\eqref{tune} holds. Since, as mentioned before, it is not clear at all how such a first step is taken in any concrete setup, we cannot say more about the reasons why the condition~\eqref{tune} should be satisfied.

\subsection{Further issues}
We would like to point out two further details that have to be addressed when considering a Lagrangian such as~\eqref{lag2}. First of all, the effective inflaton potential generated by the dominant instantons is
\begin{equation}
V_{eff}=\sum_k \tilde{\Lambda}^4\,e^{-\tilde{S}_k}\left[1-\cos\left(\vec{n}_k^{\,T}\cdot\vec{\phi}\right)\right]\,,
\end{equation}
where the classical instanton actions are given by $\tilde{S}_{k}^{(0)}=\tilde{M}\|\vec{n_k}\|$. As mentioned before, the dominant instantons are assumed, in the optimal case, to take the form $(\vec{n}_k)^i=\delta^i_k$, so that one can inflate along a diagonal long-range direction. However, as pointed out in~\cite{Montero:2015ofa}, in the absence of supersymmetry, once instantons with charges $\vec{n}_k$ and $\vec{n}_{k'}$ are included, there is generically no good reason to not include higher instanton corrections with $\tilde{S}_{k+k'}^{(0)}=\tilde{M}\|\vec{n}_{k+k'}\|$ generated by ``bound instanton states'' along the direction $\vec{n}_{k+k'}=\vec{n}_k+\vec{n}_{k'}$.\footnote{The picture is clear from the ``dual'' perspective where two charges of charge $k$ and $k'$ could form a bound state that contributes to the effective action by running in loops.} Such bound states include instantons along the diagonal direction so if, as in some examples with high kinetic mixing $\|\vec{n}_{diag}\| < \|\vec{n_k}\|$, instantons along the $\vec{n}_{diag}$ would be less suppressed than the ``elementary'' instantons, and would spoil the large range of the potential. This is another fundamental reason to consider kinetic terms such that the minimal length in charge space corresponds to the vectors $(\vec{n}_k)^i=\delta^i_k$.

Finally, let us mention another issue one needs to address if one take seriously the labels ``gravitational'' and ``gauge'' associated to the spectator instantons that satisfy the WGC and to the dominant ones that generate the inflaton potential. Since the WGC has to be satisfied in all directions in the space of axions, gravitational instantons need to generate a  (very small) mass for all the axions in the system. On the other hand, gauge instanton effects (e.g. gaugino condensation) are only expected to generate a potential for those axions that feel the gauge bosons, e.g. the axions associated with cycles wrapped by the gauge branes. One would want to use these axions to inflate. However, in string compactifications, one has generically many more axions than gauge branes. Axions associated with cycles not wrapped by gauge branes would only be stabilized by gravitational instantons, and their masses would be hierarchically smaller than those that feel the the gauge instantons. This means that the axions with an effectively trans-Planckian range would be much more massive than axions whose range is sub-Planckian. 

This situation is far from ideal. At best, both types of axions would  have masses lower than the Hubble scale in which case one would have a multi-field inflation scenario in which the lightest fields have sub-Planckian range, and only the more massive one has a trans-Planckian one. Of course, this is not the setup for simple natural inflation.  One way to avoid this would be to add as many dominant ``gauge'' instantons on top of the spectator ones as there are axions. That is, one would have to make sure that the vectors $\vec{n}_k$ span the $N$-dimensional space of charges.  This would require wrapping a stack of gauge branes around every independent cycle in the compactification manifold while satisfying various other consistency criteria; a difficult feature to engineer in detail.  

One interesting suggestion that sidesteps much of this discussion is that of \cite{Hebecker:2015rya}, which uses complex structure moduli as inflaton fields.  These do not easily dualize to the RR fields considered here and so this scenario (as far as we can tell) is not immediately eliminated by the convex hull condition.  Nevertheless, this scenario seems challenging to implement in detail.  It would be interesting to understand further if a more generalized version of the WGC could place constraints on this class of models.

\section{Conclusions}
In this note we have demonstrated that strong-WGC prohibits large field ranges, independently of the proxy used to define the inflaton range.  Thus, the WGC cannot accommodate non-monodromic natural inflation.  Using the loophole to evade this constraint requires one to overcome several difficult hurdles. It is an interesting possibility which deserves further study, but so far there is no compelling evidence that the strong-WGC can be violated, nor how it's violation may be exploited in a concrete framework.

\subsection*{Acknowledgments}
We would like to thank
 Thomas Bachlechner, Pablo Camara, Cody Long, Miguel Montero, Thomas van Riet, Angel Uranga and Irene Valenzuela
  for useful discussions.  This work is supported in part by the DOE grant DE-FG-02-95ER40896 and the HKRGC grants 
HUKST4/CRF/13G, 604231 and 16304414.

\bibliography{WGC-again}\bibliographystyle{utphys}

\end{document}